\documentclass[%
 reprint,
 amsmath,amssymb,
 aps,
]{revtex4-1}
\usepackage{filecontents}
\usepackage{siunitx} 
\usepackage{multirow} 
\usepackage[T1]{fontenc}
\usepackage{graphicx}
\usepackage{dcolumn}
\usepackage{bm}

\begin{document}


\title{Temperature and electric field dependence of spin relaxation in graphene on SrTiO$_3$}
\author{S. Chen}
\email{s.chen@rug.nl}
\affiliation{%
Zernike Institute for Advanced Materials, University of Groningen, Nijenborgh 4, 9747 AG Groningen, The Netherlands\\
}
\author{R. Ruiter}%
\affiliation{%
Zernike Institute for Advanced Materials, University of Groningen, Nijenborgh 4, 9747 AG Groningen, The Netherlands\\
}
\author{V. Mathkar}
\affiliation{%
Zernike Institute for Advanced Materials, University of Groningen, Nijenborgh 4, 9747 AG Groningen, The Netherlands\\
}
\author{B. J. van Wees}
\affiliation{%
Zernike Institute for Advanced Materials, University of Groningen, Nijenborgh 4, 9747 AG Groningen, The Netherlands\\
}
\author{T. Banerjee}
 \email{t.banerjee@rug.nl}
\affiliation{%
Zernike Institute for Advanced Materials, University of Groningen, Nijenborgh 4, 9747 AG Groningen, The Netherlands\\
}

\date{\today}

\begin{abstract}
The theoretically predicted intrinsic spin relaxation time of up to \SI{1}{\micro\second} in graphene along with extremely high mobilities makes it a promising material in spintronics. Numerous experimental studies, however, find the spin lifetime in graphene to be several orders of magnitude below that theoretically predicted. Additionally, analyses of the spin relaxations mechanisms in graphene using conventional processes such as Elliot-Yaffet and D'Yakonov-Perel' show a coexistence of both, with no clear dominance. Central to these experimental discrepancies is the role of the local environment including that of the underlying substrate. In this work, we use the electronically rich platform of SrTiO$_3$ with broken inversion symmetry and study spin transport in graphene in the presence of surface electric fields. 
We find spin relaxation time and length as large as $0.96\pm$ \SI{0.03}{\nano\second} and $4.1\pm$ \SI{0.1}{\micro\meter}, respectively at \SI{290} {\kelvin} in graphene, using non-local spin valve studies and find a non monotonous dependence with temperature, unlike that observed in other substrates. Analysis of the temperature dependence indicate the role of surface electric dipoles and electric field driven electronic and structural phase transitions unique to SrTiO$_3$, to spin transport and spin relaxation in graphene.  

\begin{description}
\item[PACS numbers]
\end{description}
\end{abstract}

\maketitle


\begin{figure*}[h]
    \centering
    \includegraphics[width=\linewidth]{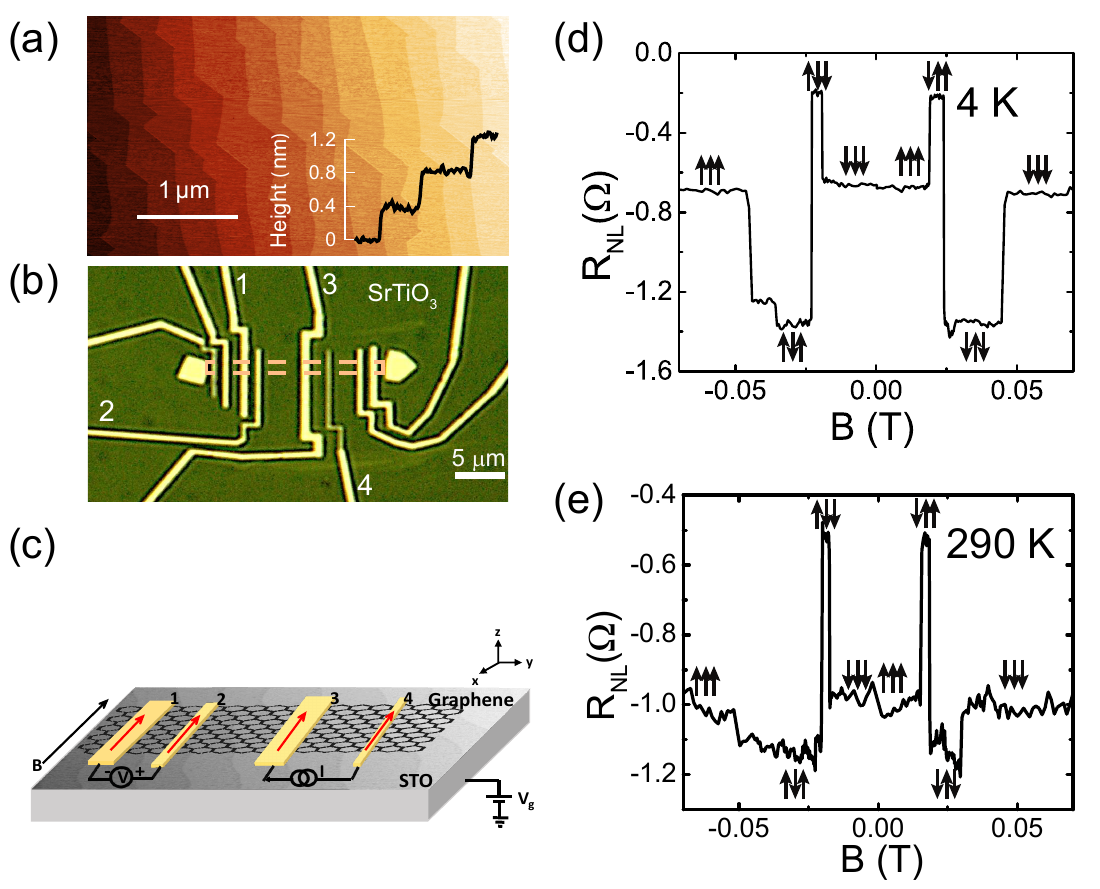}
    \caption{(a) Atomic force microscope image of a TiO$_2$ terminated STO substrate after surface treatment and annealing at \SI{960}{\celsius}. The inset shows a height profile of the terrace steps. (b) Optical microscope image of the device. Co/AlO$_x$ contacts are in yellow and the graphene flake is indicated by the orange dash line. (c)  Layout of the measurement circuit, where the Co/AlO$_x$ electrodes are in gold. A current is injected through contact 3 and detected by contact 1-2. (d,e) Measurements of non-local spin valve at \SI{4} {\kelvin} (top) \SI{290} {\kelvin} (bottom). Magnetic switching corresponding to contacts 1, 3, 2 are observed.}
    \label{fig:device}
\end{figure*}

Charge conduction and spin transport parameters in two-dimensional graphene are strongly influenced by extrinsic factors related to their local environment. Extrinsic influences range from the specifics of the underlying substrate (suspended, encapsulated or high dielectric constant),\cite{ertler_electron_2009,guimaraes_controlling_2014} the quality of the contacts \cite{han_tunneling_2010,maassen_contact-induced_2012} to spin-orbit effects due to adatoms. \cite{Hernando2009,Fratini2013,Gurram2018}
Despite significant improvements either on enhancing the graphene quality including encapsulation on an atomically flat two dimensional hexagonal Boron Nitride (hBN) substrate or by resolving extrinsic influences, the experimentally measured spin lifetime in graphene is orders of magnitude smaller than theoretically predicted.  \cite{zomer_long-distance_2012,Dean2010,Pep2015, Kamalakar2015,Dlubak2012,Avsar2011} Furthermore, conventional spin relaxation mechanisms such as Elliot-Yaffet and D'Yakonov-Perel' fail to unambiguously explain the nature and dominance of the spin dephasing processes in graphene on different substrates.\cite{zomer_long-distance_2012,Hernando2009,Fratini2013,Gurram2018,Roche2014} Spin dephasing can originate from a multitude of effects such as flexural distortions, ripples, local magnetic moments, to name a few, but understanding of their precise micoroscopic mechanism still remains elusive. \cite{Hernando2009,Fratini2013,Gurram2018}

In this context SrTiO$_3$ (STO) lends itself as an interesting choice of substrate to study spin relaxation mechanisms in graphene.\cite{Levy2014} STO has an atomically flat surface, similar to that of hBN, with roughness of 90 -150 pm and no dangling bonds. However, unlike hBN, STO is electronically versatile. This stems from the remarkably large dielectric constant ($\varepsilon_\mathrm{r}$) of 300 at room temperature that increases non-linearly to $>20,000$ at \SI{4} {\kelvin}.\cite{sakudo_dielectric_1971} Further, distinct from most other substrates on which charge and spin transport in graphene has been studied, the broken inversion symmetry at the surface of STO leads to Rashba spin orbit fields that can be tuned by an electric field. \cite{kamerbeek_electric_2015} Recently it was demonstrated that electric dipoles, formed at the surface of STO, results in a large out of plane electric polarization that influences the charge transport in graphene. \cite{Sahoo2018}\ STO undergoes a ferroelastic transition changing from cubic ($a$ = \SI{3.905}{\angstrom}) to tetragonal symmetry ($c/a = 1.0056$) at $T$ = \SI{105}{\kelvin}.\cite{lytle_xray_1964} This is accompanied by structural domains that can be moved with an external gate-bias.\cite{honig_local_2013,Frenkel2017} The movement of such structural ferroelastic domains at low temperatures can lead to modulations in the surface potential in STO that causes local fluctuations in the carrier density of graphene. Thus STO offers an electronically rich transport platform for graphene-based devices. Recent studies on the charge transport in graphene on STO \cite{DasSarma2012,couto_transport_2011,couto_random_2014,sachs_ferroelectric-like_2014,saha_unconventional_2014,Kang2017,Ryu2017} discusses the influence of the high $\epsilon_\mathrm{r}$ and its role in screening impurities and improving the charge mobility, $\mu$, in graphene.\cite{adam_self-consistent_2007}\ 

\begin{figure*}[h]
\centering
\includegraphics[width=\linewidth]{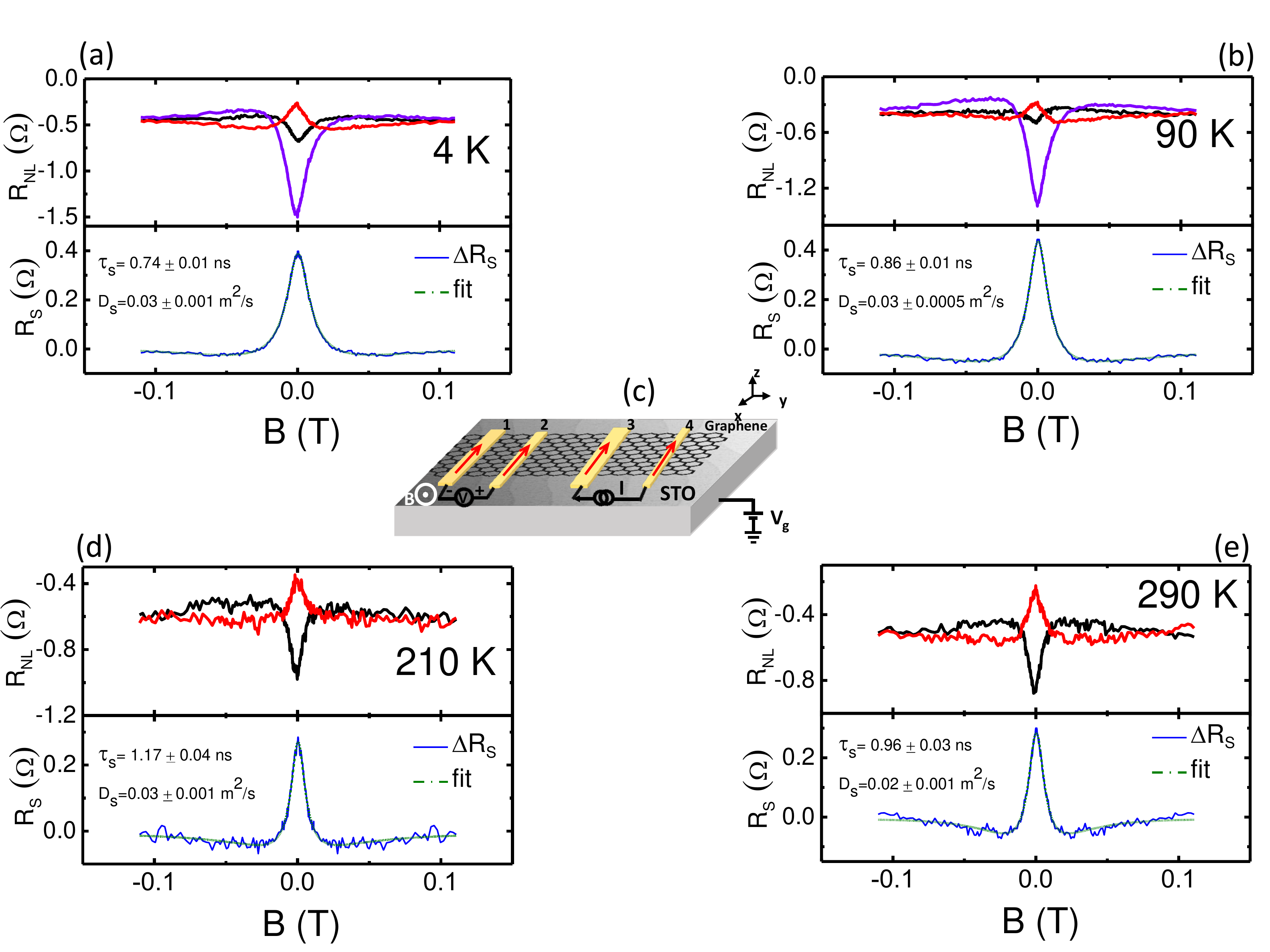}
\caption{(a, b, d, e) Hanle measurements of $\downarrow\uparrow\uparrow$ (black) and $\uparrow\uparrow\uparrow$ (red) $\downarrow\uparrow\downarrow$ (purple) configurations of contact 1, 2, 3 respectively and the calculated Hanle signal (blue, bottom panel) after the background subtraction and fitting with the steady state solution to the Bloch equation in the diffusive regime. Shown are for \SI{4} {\kelvin} (a), \SI{90} {\kelvin} (b), \SI{210} {\kelvin} (d) and \SI{290} {\kelvin} (e).  Error bars are derived from fitting errors. (c) Schematic of Hanle measurements with magnetic field out of plane.}
\label{fig:T-dep_dev1}
\end{figure*}

Interestingly, in spite of the above studies, the influence of temperature and electric field driven structural and electronic phase transition in STO, as well as the large intrinsic Rashba spin orbit fields, on spin transport in graphene is largely unexplored. In this work, we study spin transport in graphene on STO for the first time and investigate the effects of temperature and electric field using spin injection contacts of Co/AlO$_x$. We find the spin relaxation time at \SI{290} {\kelvin} to be as long as $0.96 \pm$ \SI{0.03}{\nano\second}, with a spin relaxation length of $4.1 \pm$ \SI{0.1}{\micro\meter}. To investigate the role of the surface dipoles and the large, temperature dependent, non-linear dielectric constant on spin transport in graphene and across the ferroelastic transition in STO, spin transport measurements are performed at different temperatures. A non monotonous temperature dependence of spin transport parameters, characterized by spin lifetime and diffusion constant  is observed. We find the spin transport parameters to be lower at \SI{4}{\kelvin} than at \SI{290} {\kelvin}- contrary to that expected and an observation not reported earlier using other substrates such as SiC or SiO$_2$. Furthermore, we find that the gate dependence of the spin relaxation parameters at \SI{4}{\kelvin} is associated with the modulation of the strength of surface dipoles in STO. An analysis of the spin relaxation mechanisms reveals the coexistence of both Elliot-Yaffet and D'Yakonov-Perel' scattering processes.\

To investigate spin transport in graphene on STO, lateral spin valves of exfoliated graphene on TiO$_2$ terminated STO were fabricated.  One side polished STO (100) substrates (Crystec GmbH) were treated with a standard protocol \cite{kawasaki_atomic_1994,koster_quasi-ideal_1998} to achieve a TiO$_2$ terminated surface. An atomic force microscope (AFM) scan of one such  terminated STO surface is shown in figure \ref{fig:device}a.  Graphite (grade ZYA) was exfoliated on a clean SiO$_2$/Si wafer and single layer graphene was selected based on optical contrast. These flakes were transferred from the SiO$_2$ to the desired area on the STO substrate using a polycarbonate dry pick-up technique.\cite{zomer_fast_2014} Polycarbonate residues left behind after the transfer were removed by treating the substrate at \SI{50}{\celsius} with chloroform for a period of several hours to days. Electrical contacts were defined using electron beam lithography and deposited using electron beam evaporation in multiple steps. A tunnel barrier was deposited in a two-step process: first \SI{0.4} {\nano\meter} of aluminium was deposited and oxidized for 10 minutes in a pure oxygen atmosphere. This step was repeated once more to obtain a $\approx$\SI{1} {\nano\meter} thick AlO$_x$ tunnel barrier. Thereafter \SI{35} {\nano\meter} of ferromagnetic cobalt was deposited and capped with \SI{5} {\nano\meter} aluminum layer to prevent cobalt from oxidizing. The substrate was bonded on a chip carrier using silver paste, which serves as the back gate during our transport measurements. An optical image of the device is shown in figure \ref{fig:device}b.\

\begin{figure*}[h]
    \centering
    \includegraphics[width=\linewidth]{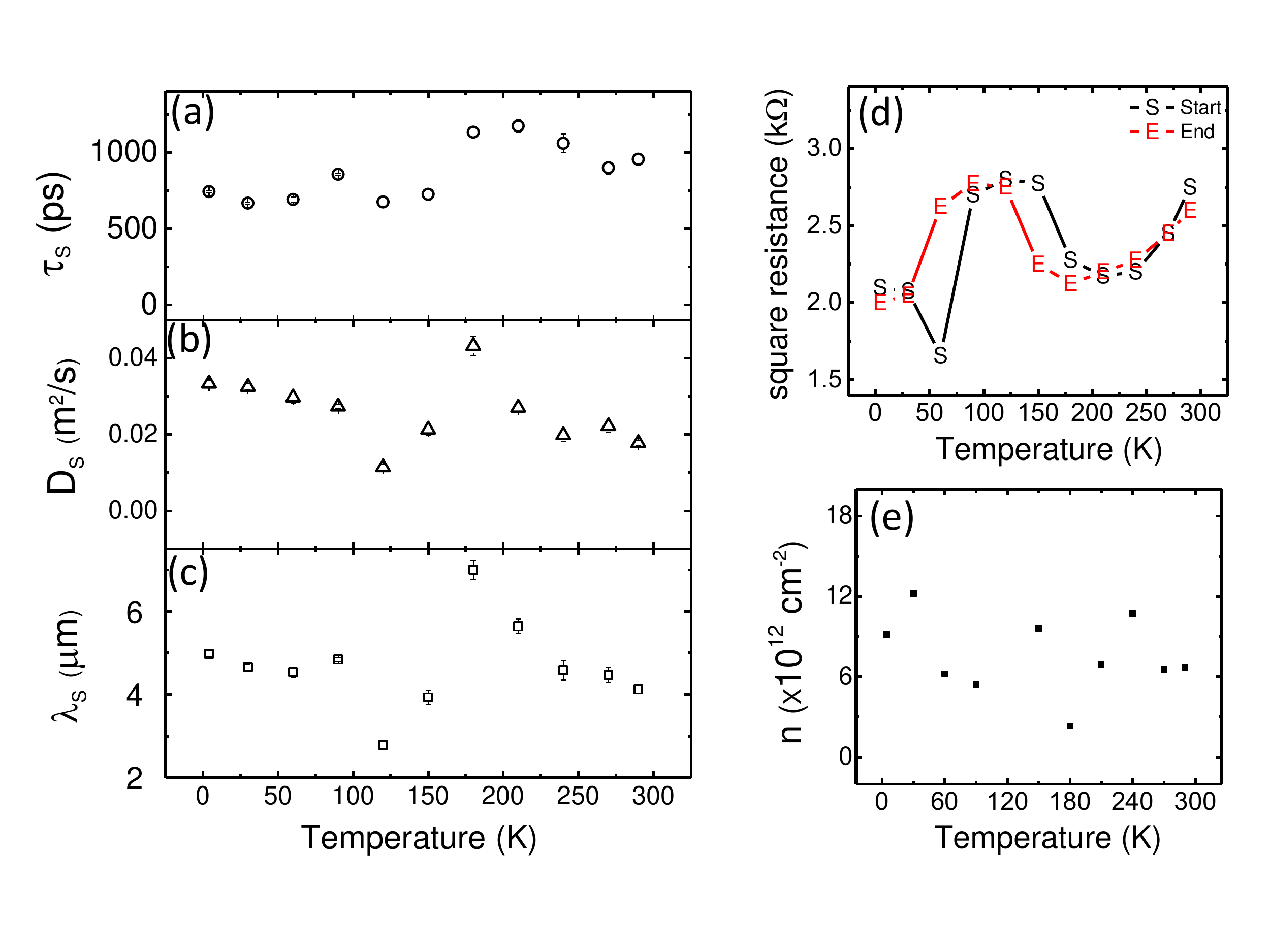}
    \caption{(a) The extracted spin relaxation time $\tau_\mathrm{s}$, (b) spin diffusion constant $D_\mathrm{s}$ and (c) the calculated spin relaxation length $\lambda_\mathrm{s} = \sqrt{D_\mathrm{s}\tau_\mathrm{s}}$ as a function of temperature from Hanle measurements. Error bars are derived from fitting errors. (d) Square resistance of the graphene channel used for the Hanle measurements shown in figure \ref{fig:T-dep_dev1}b. The square resistance was measured at the start and at the end of a measurement sequence for each measurement temperature.(e) Carrier density versus temperature calculated from spin diffusion constant $D_s$ and square resistance $R_{sq}$ of graphene}
    \label{fig:R-sq_Hanle}
\end{figure*}

Spin transport measurements were performed on the device, using a non-local geometry. This geometry, shown in figure \ref{fig:device}c, separates the charge current path from the voltage contacts thus excluding spurious signals. The center to center separation between contacts 2 and 3 is \SI{4.8} {\micro\meter}. In this configuration both spin valve as well as Hanle precession measurements can be performed. Assuming $\varepsilon_\mathrm{r}$ in STO to be 24000, we calculate the mobility $\mu$ of the sample to be 658 $\mathrm{cm}^{2}/\mathrm{Vs}$ at 4 K obtained using $\mu=1/e d\sigma/dn$. Spin valve measurements were performed by sweeping an in-plane magnetic field $B$ in the $y$-direction and measuring the non-local resistance $R_\mathrm{NL}=V/I$, using lock-in techniques with frequencies $<\SI{15}{\hertz}$. \cite{popinciuc_electronic_2009} Figures \ref{fig:device}d, e show the response of one of the non-local spin valves at \SI{4} and \SI{290} {\kelvin} respectively. Three clear switches are present at both temperatures, corresponding to contacts 1, 3, and 2. \ 

Additionally temperature dependent Hanle precession measurements were performed and shown in figures \ref{fig:T-dep_dev1}a,b,d,e, are for different temperatures and magnetic configurations of the electrodes $\downarrow\uparrow\uparrow$, $\uparrow\uparrow\uparrow$ and $\downarrow\uparrow\downarrow$ for the contacts 1, 2, 3 ($\uparrow$, $\downarrow$ refers to the magnetization of the electrodes in y and -y direction respectively). The $B$-field is swept out-of-plane ($z$-direction) while measuring $R_\mathrm{NL}$. Spins injected at contact 3 will start to precess around the $B$-field, thereby changing the projected spin component along the $y$-direction where they are detected by contacts 1-2. The resultant Hanle curves of $\downarrow\uparrow\uparrow$, $\uparrow\uparrow\uparrow$ and $\downarrow\uparrow\downarrow$ configurations are shown in the top panels in figures \ref{fig:T-dep_dev1}a,b,d,e (black, red and purple curves respectively). A common background is subtracted to obtain the pure spin signal using: $R_\mathrm{s}=1/2 \left( R_{\downarrow\uparrow\uparrow} - R_{\uparrow\uparrow\uparrow} \right)$ as detailed in ref. \cite{zomer_long-distance_2012}. The blue curve in the bottom panel in figures \ref{fig:T-dep_dev1}a,b,d,e is the data after background subtraction. Thereafter, it was fitted with the steady state solution to the Bloch equation in the diffusive regime.\cite{popinciuc_electronic_2009,fabian_semiconductor_2007} From this, the spin diffusion constant $D_\mathrm{s}$ and the spin relaxation time $\tau_\mathrm{s}$ were obtained and the spin relaxation length was calculated using: $\lambda_\mathrm{s} = \sqrt{ D_\mathrm{s}\tau_\mathrm{s} }$. From the fitting, we find $\tau_\mathrm{s}= 0.96 \pm$ \SI{0.03} {\nano\second}, $D_\mathrm{s}= 0.02 \pm$ \SI{0.001} {cm^2/s} and $\lambda_\mathrm{s}=4.1 \pm$ \SI{0.1} {\micro\meter} at \SI{290} {\kelvin}. The extracted values at other temperatures can be found in the bottom panel of figures \ref{fig:T-dep_dev1}a,b,e. We find that the value at \SI{290} {\kelvin} compares well to that reported on hBN substrate\cite{zomer_long-distance_2012}, however temperature dependence studies were not reported by those authors. Our findings establishes the effectiveness of STO as a suitable platform to study spin transport in graphene on STO.

In order to investigate the influence of temperature driven electronic and structural phase transitions in STO on spin transport in graphene, we exploit the temperature dependent Hanle measurements. Before starting the spin transport measurements, a back gate was swept between \SI{0}{\volt} $\sim$ \SI{-70} {\volt} $\sim$ \SI{+70} {\volt} $\sim$ \SI{0} {\volt} at \SI{4} {\K} in order to characterize the charge transport properties in graphene on STO (see Figure S2, Supporting Information). Thereafter spin transport measurements were carried out while heating up the device. Typically the measurements were recorded after $R_\mathrm{sq}$, the square resistance of graphene, stabilized over time. The measurements were performed using the non-local geometry (figure \ref{fig:T-dep_dev1}c) in a temperature range between $4$ and \SI{290} {\kelvin}. A non monotonous variation of the spin relaxation time, $\tau_\mathrm{s}$, with temperature is observed up to \SI{180} {\K} which decreases thereafter with increasing temperature as shown in (figure \ref{fig:R-sq_Hanle}a-c )The variation of the spin diffusion constant, $D_\mathrm{s}$, with temperature is different from that of the spin relaxation time as shown in figure \ref{fig:R-sq_Hanle}b. The calculated value of the spin relaxation length, $\lambda_\mathrm{s}$ shows a maximum \SI{7} {\micro\meter} at \SI{180} {\kelvin}. We have observed similar trends in the temperature dependence of spin transport, but with a shorter spin relaxation time, on devices fabricated on other STO substrates.\ 

The spin transport parameters are usually known to decrease with increasing temperature, due to electron-phonon scattering. Temperature dependent studies of spin transport parameters in graphene are scarce and two earlier studies\cite{maassen_contact-induced_2012,han_tunneling_2010} on SiC and Si substrate report a decrease of spin transport parameters with increasing temperature. This is ascribed to the enhanced electron-phonon scattering at higher temperatures. Although we find a similar trend for temperatures above \SI{180} {\K} in our devices, the variation at lower temperatures is contrary to expectations. To understand this variation, we first look into the temperature dependence of $R_\mathrm{sq}$ of graphene (figure \ref{fig:R-sq_Hanle}d). We note that the spin transport parameters are unchanged in the temperature regime \SI{4} {\K}-\SI{180} {\K}, where the variation of $R_\mathrm{sq}$ is the largest. This is clearer if we analyse the carrier density using Einstein relation
$\sigma=e^2\nu D_\mathrm{c}$, assuming $D_\mathrm{c}\approx D_\mathrm{s}$:
\begin{equation}
    \label{eq:Rsq_n}
    n=\frac{(\hbar v_\mathrm{F})^2\pi}{g_\mathrm{s}g_\mathrm{v}D_\mathrm{s}^2 R_\mathrm{sq}^2 e^4}
\end{equation}
where $R_\mathrm{sq}$ is the square resistance of graphene, $v_\mathrm{F}$ is the Fermi velocity, $\hbar$ is the reduced Planck's constant, $g_\mathrm{s}$ = 2, $g_\mathrm{v}$ = 2 are the spin and valley degeneracy respectively, $D_\mathrm{c}$ is the charge diffusion coefficient and $\nu$ is the density of states at the Fermi energy. The calculated carrier density varies between $3\times10^{12} \mathrm{cm}^{-2}$-$1\times 10^{13} \mathrm{cm}^{-2}$ (figure \ref{fig:R-sq_Hanle}e) in this temperature range and is consistent with other similarly fabricated devices in a Hall bar geometry ($10^{12}  \mathrm{cm}^{-2}- 10^{13}  \mathrm{cm}^{-2}$). The observed fluctuation of the carrier density with temperature is mainly attributed to the uncertainty in the determination of $D_\mathrm{s}$. The determination of $D_\mathrm{s}$ is sensitive to the detailed structure of the Hanle curve and is much more sensitive than the determination of $\tau_\mathrm{s}$.  In equation 1, $n$ varies as $\frac{1}{D_\mathrm{s}^2}$ and $D_\mathrm{s}$ is obtained from the fitting of the Hanle data. Any uncertainty in the determination of $D_\mathrm{s}$ is thus amplified in the calculation of $n$. Given the uncertainty in the fitting procedure, a temperature independence of the carrier density cannot be strictly excluded. The important point however is, that despite the fluctuations in $n$, graphene is in a high carrier density regime where changes in carrier density do not have a big influence on the spin relaxation in graphene.\

\begin{figure*}[h]
    \centering
    \includegraphics[width=\linewidth]{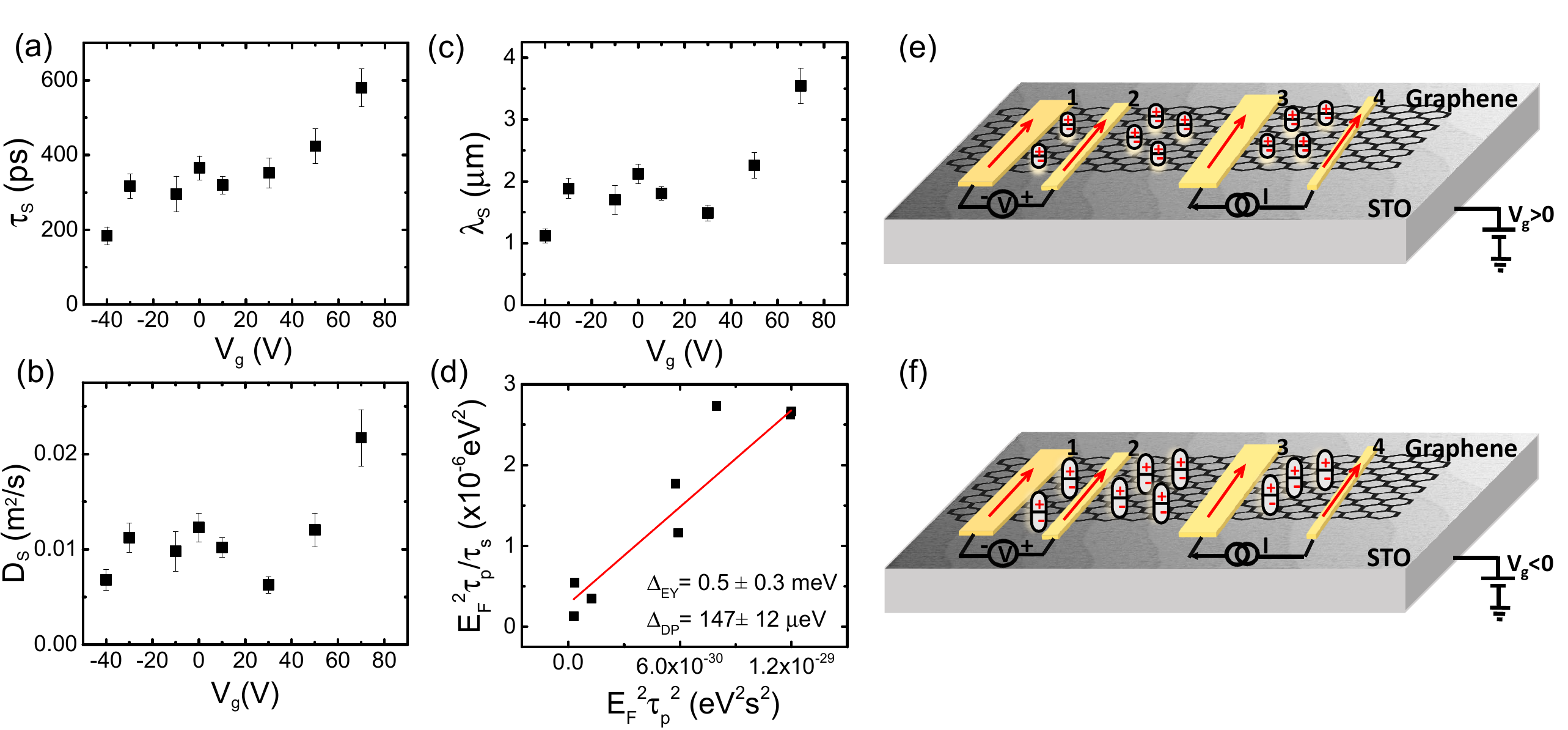}
    \caption{Gate dependent spin transport parameters at \SI{4}{\kelvin}. (a) (b) (c) gate dependence of $\tau_\mathrm{s}$, $D_\mathrm{s}$ and $\lambda_\mathrm{s}$. The gate is swept between \SI{+70}{\volt} to \SI{-70}{\volt}. Higher $\tau_\mathrm{s}$, $D_\mathrm{s}$ and $\lambda_\mathrm{s}$ values are observed at positive positive gate voltage. Note: below \SI{-40}{\volt}, the Hanle curves cannot be fitted with the steady state solution to the one-dimensional Bloch equation. (d) $\frac{E_F^2\tau_p}{\tau_s}$ versus $E_F^2\tau_p^2$ at \SI{4}{\kelvin}. The solid line is the fit using Equation \ref{eq:EY_DP}. The error bars are the fitting errors. (e)(f) Surface dipoles at $V_\mathrm{g} > 0$ and $V_\mathrm{g} < 0$. At positive gate voltage, the surface dipoles are suppressed by the electric field; at negative gate voltage, the surface dipoles are enhanced.
    }
    \label{fig:Vg-dep_tau}
\end{figure*}

To understand the contribution of spin absorption to the observed variation of the spin lifetime, we next discuss the invasiveness of the contacts with temperature. The injected spins from the low resistive ferromagnetic contacts to the high resistive channel can be backscattered into the electrodes depending on the ratio between the resistance of the contacts and the spin transport channel. Thus a change in either the contact resistance ($R_\mathrm{c}$) or in the channel resistance ($R_\mathrm{sq}$), measured in a three and four probe measurement geometry, can lead to a change in the extracted spin transport parameters. In our case, we measure contact resistances of \SIlist{8;21;22;13}{\kohm} for contacts 1, 2, 3, 4 respectively and find that both the contact resistance and square resistance have very little fluctuation with temperature. To further quantify this, the invasiveness of the contacts is evaluated using the R parameter, $R=(R_\mathrm{c}/R_\mathrm{sq})W$, where W is the width of the graphene channel.\cite{maassen_contact-induced_2012} $R/\lambda_\mathrm{s}$ varies from $1.1$ to $3.5$ (for $L/\lambda_\mathrm{s} = 0.9 \sim 1.6$) and as discussed in ref. \cite{maassen_contact-induced_2012}, such a fluctuation in  $R/\lambda_\mathrm{s}$ can maximally contribute to a change of \SI{20}{\%} in the spin transport parameters, (shown in figure S3 in supplementary information). Thus it can be inferred that the invasiveness of the contacts do not play a crucial role in the observed temperature variation of the spin transport parameters.\

\begin{figure*}[h]
    \centering
    \includegraphics[width=\linewidth]{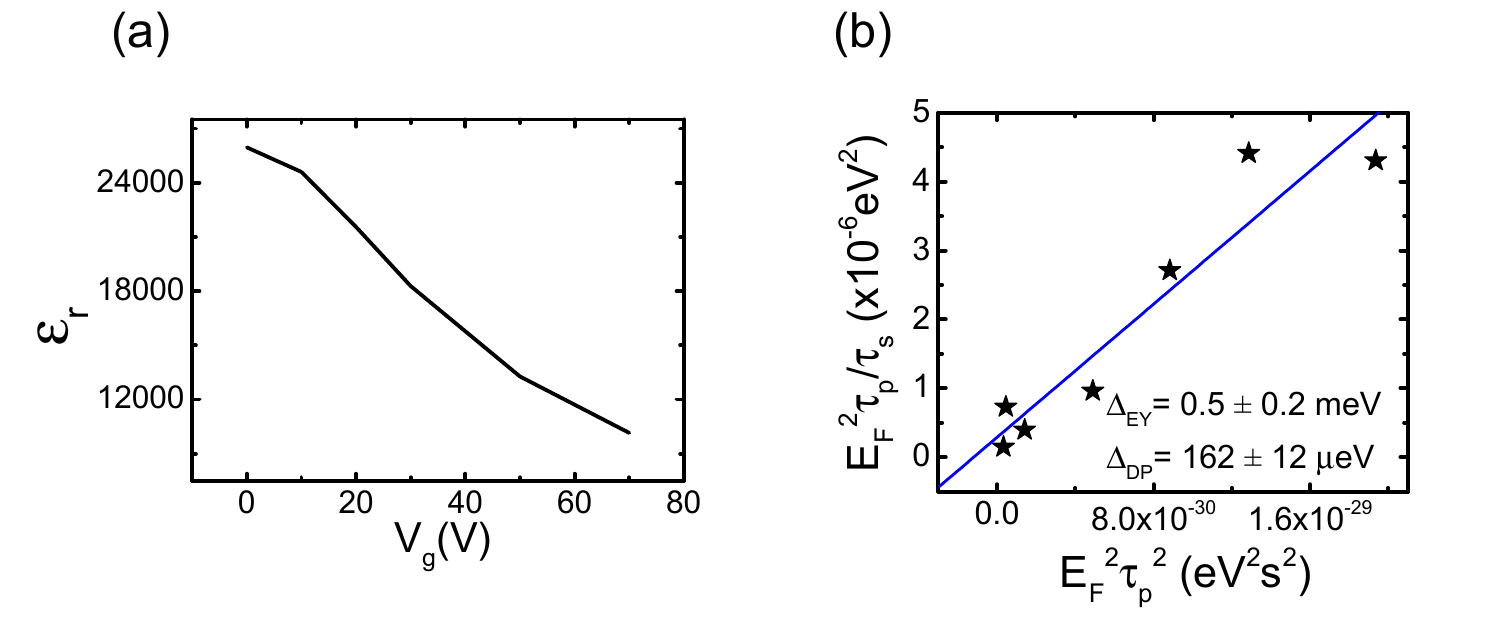}
    \caption{ (a) Gate dependent $\varepsilon_\mathrm{r}$ at 4 K calculated using Equation \ref{eq:epsilon}. (b) $\frac{E_F^2\tau_p}{\tau_s}$ versus $E_F^2\tau_p^2$ at \SI{4}{\kelvin} taking into account the variation of $\varepsilon_\mathrm{r}$ with gate bias. The calculated data points incorporating the dielectric constant correction is represented by the stars and the blue solid line is the fit using Equation \ref{eq:EY_DP}. The error bars are the fitting errors.}
    \label{fig:epsilonwithgate}
\end{figure*}

We now consider the role of spin-orbit coupling in STO to explain the reduction in spin relaxation times at low temperatures. An electric field induced (via spin orbit coupling) change in the spin relaxation time has been demonstrated earlier in graphene and also in doped-STO.\cite{guimaraes_controlling_2014,kamerbeek_electric_2015} In our case, the origin of the electric field is intrinsic to the STO substrate. The surface of STO has a broken inversion symmetry and is found to harbor surface dipoles where the oxygen atoms are displaced outwards.\cite{sachs_ferroelectric-like_2014,heifets_semi-empirical_2000} The strength of these surface dipole moments is calculated to be $P$ = \SI{-13.89}{\micro C/cm^2} which increases to $P$ = \SI{-34.90} {\micro C/cm^2} with the graphene layer. \cite{Sahoo2018,Vanderbilt} This electric field, pointing outwards from the substrate plane, originates from the surface dipoles in STO. The strength of this electric field is enhanced by the increased $\varepsilon_\mathrm{r}$ in STO at low temperatures. This intrinsic electric field effectively influences the spin relaxation time in graphene via spin-orbit coupling. With increasing temperature, a reduction in the surface electric field occurs due to the slowly relaxing surface dipoles \cite{sachs_ferroelectric-like_2014}, leading to an increase in the spin relaxation time. At temperatures above \SI{180} {\K}, the electron phonon scattering plays an important role, leading to a slight decrease of $\tau_\mathrm{s}$, $D_\mathrm{s}$ and $\lambda_\mathrm{s}$.\ 

Further, structural transition in STO at 105 K, that leads to the formation of long striped domains and induces ripples on the graphene surface, are important considerations in the analysis of the temperature dependence of  transport in graphene. Recent studies show that below the cubic to tetragonal phase transition temperature of \SI{105} {\kelvin}, differently oriented large (micron size) tetragonal domains are formed in STO where the local electrostatics are different than at the domain walls. \cite{honig_local_2013,Frenkel2017} These will induce surface potential modulations across the graphene sheet (see the charge transport data in figure S2, Supporting Information) on STO and will vary strongly with temperature. The induced rumplings will act as local scattering centers at low temperatures leading to enhanced spin dephasing and hindering efficient spin transport.\

From our detailed analysis above, we infer that the variation of the carrier density and contact resistance with temperature can be eliminated as factors influencing the temperature dependence of the spin transport in graphene. However, as discussed above, mechanisms such as electric field induced by the surface dipoles, intrinsic spin orbit coupling and potential modulations due to temperature induced rippling at the graphene interface all play a cumulative role in the observed non monotonous temperature variation of the spin transport parameters.\ 

Using STO as a back gate, we can tune the surface dipoles on the surface of STO at 4 K as shown in figure \ref{fig:Vg-dep_tau}. $\varepsilon_\mathrm{r}$ of STO increases from 300 at room temperature to more than 24,000 at 4 K. A back gate modulation of the surface dipoles through a 0.5 mm thick STO substrate is realized at 4 K. Beyond 4 K, the decreased $\varepsilon_\mathrm{r}$ render the back gating less efficient, making it difficult to draw any reasonable conclusions on the gate dependence of spin relaxation at higher temperatures. We have swept the back gate between 70 V to -70 V and wait for 600 s before the start of each spin transport measurement. As shown in figure \ref{fig:Vg-dep_tau}e and f, the intrinsic surface dipole strength will be enhanced at negative gate voltage and reduced at positive gate voltage, thus influencing the spin relaxation time. We observe that $\tau_\mathrm{s}$, $D_\mathrm{s}$ and $\lambda_\mathrm{s}$ increases up to a factor of two at higher positive $V_\mathrm{g}$, while at negative and small positive gate voltage, these parameters do not seem to change much, as shown in figures \ref{fig:Vg-dep_tau}a-c.\

There are two prevalent spin relaxation mechanisms in graphene. The first is the Elliot-Yafet (EY) mechanism, where the spin loses its direction by scattering with the impurities and the spin-relaxation time is proportional to the momentum relaxation time. The second mechanism is the D'yakonov-Perel'(DP) mechanism, where the spin precesses in a spin-orbit field between two momentum scattering events and the spin relaxation time is inversely proportional to the momentum relaxation time. To further quantify the relative contribution of the Elliot-Yafet (EY) and D'yakonov-Perel'(DP) mechanisms in our case, we assume $D_\mathrm{s} \approx D_\mathrm{c}$ and calculate the momentum relaxation time $\tau_\mathrm{p}$ according to $D \sim {v_F}^2\tau_\mathrm{p}$. Following the standard analysis as used by Zomer \emph{et al.}, Jo \emph{et al.} and Gurram \emph{et al.}\cite{zomer_long-distance_2012,Jo2011,Gurram2018}, we analyze the relation between $\tau_\mathrm{p}$ and $\tau_\mathrm{s}$ using the equation:

\begin{equation}
    \label{eq:EY_DP}
        \frac{{E_\mathrm{F}}^2\tau_\mathrm{p}}{\tau_\mathrm{s}}=\Delta_\mathrm{EY}^2+(\frac{4\Delta_\mathrm{DP}^2}{\hbar^2}){E_\mathrm{F}}^2\tau_\mathrm{p}^2
\end{equation}
where $E_\mathrm{F}$ is the Fermi energy, $\Delta_{EY}$ and $\Delta_{DP}$ are the spin orbit coupling for EY and DP mechanism respectively. Figure \ref{fig:Vg-dep_tau}d shows the $\frac{E_\mathrm{F}^2\tau_p}{\tau_s}$ versus $E_\mathrm{F}^2\tau_p^2$ dependence. From the fitting, we extract $\Delta_{EY}$ = \SI{532}{\mu eV} and $\Delta_{DP}$ = \SI{147}{\mu eV}. \

We further consider the variation of $\varepsilon_\mathrm{r}$ by an applied gate bias and analyze the spin relaxation mechanism in graphene. $\varepsilon_\mathrm{r}$ of STO can be calculated using:\cite{Kamerbeek2015,Suzuki1997}

\begin{equation}
    \label{eq:epsilon}
    \varepsilon_\mathrm{r}(T,E)=\frac{b(T)}{\sqrt{a(T)+E^2}}
\end{equation}

where $T$ is the temperature, $E$ is the electric field, fitting parameters $b(T) = 1.37\times10^{7}+4.29\times10^{7}(\frac{T}{100})$V/cm and $a(T)=[b(T)/\varepsilon_\mathrm{r}(T,0)]^2 \mathrm{V}^2/\mathrm{cm}^2$, and $\varepsilon_\mathrm{r}(T,0)$ is expressed using Barrett's formula:\cite{Barrett1952}

\begin{equation}
    \label{eq:epsilon0}
    \varepsilon_\mathrm{r}(T,0)=\frac{1635}{coth(44.1/T)-0.937}
\end{equation}

 The calculated variation of $\varepsilon_\mathrm{r}$ on back gate is shown in figure \ref{fig:epsilonwithgate}a. We incorporate the variation of $\varepsilon_r$ in our calculation of $E_\mathrm{F}$ using  $n=4{E_\mathrm{F}}^2/(g_\mathrm{s}g_\mathrm{v}\pi\hbar^2v_\mathrm{F}^2)=\frac{\varepsilon_\mathrm{r}\varepsilon_0}{et}\Delta V_\mathrm{g}$, where $\varepsilon_0$ is the vacuum permittivity and t is the thickness of STO (0.5 mm). Figure \ref{fig:Vg-dep_tau}d is now reanalyzed considering this variation of $\varepsilon_\mathrm{r}$ by the applied electric field and this is shown in figure \ref{fig:epsilonwithgate}b. From figure \ref{fig:epsilonwithgate} b, we obtain $\Delta_{EY}$ = \SI{537}{\mu eV} and $\Delta_{DP}$ = \SI{162}{\mu eV} (the respective numbers in figure \ref{fig:Vg-dep_tau}d are $\Delta_{EY}$ = \SI{532}{\mu eV} and $\Delta_{DP}$ = \SI{147}{\mu eV}). Reported values of spin relaxation mechanisms in graphene on hBN and on other substrates\cite{zomer_long-distance_2012,Jo2011,Gurram2018} shows $\Delta_{DP}$ to be varying between 40-200 $\mu$eV and $\Delta_{EY}$ between 0.5 meV and 2.3 meV with no clear dominance of either mechanism. The values we obtain from our fittings are similar to these values. Spin relaxation rates are further analyzed: $\tau_\mathrm{(s,EY)}$ = 0.2-5.6 $\mathrm{ns}^{-1}$ and $\tau_\mathrm{(s,DP)}$ = 0.02-0.09 $\mathrm{ns}^{-1}$, and are found to be of the same order for EY and DP spin relaxation mechanisms. We conclude that for graphene on STO, there is a coexistence of both mechanisms with no clear dominance of one mechanism over the other as reported for other substrates. \cite{zomer_long-distance_2012,Jo2011,Gurram2018}\
 
We report on the first observation of spin transport in graphene on TiO$_2$ terminated STO with broken inversion symmetry. A spin relaxation time and length of $0.96 \pm $\SI{0.03} {\nano\second} and $4.1 \pm$ \SI{0.1} {\micro\meter} is obtained at 290 K alongwith a non monotonous variation of the spin transport parameters at low temperatures. Our work shows that spin transport in graphene on STO is influenced by the cumulative effect of surface electric dipoles, intrinsic spin orbit coupling and temperature induced rippling of the graphene interface. Gate dependence of the spin relaxation parameters at 4 K is attributed to the modulation of the strength of the surface dipoles in STO, while an analysis of the spin relaxation mechanism shows the coexistence of both EY and DP scattering processes. Our studies on integrating graphene with complex oxides opens new opportunities to study proximity induced functionalities at such interfaces, useful for future spintronics and optoelectronics applications.

We acknowledge J. G. Holstein, H. M. de Roosz, H. Adema and T. J. Schouten for their technical support and S. Omar, A. M. Kamerbeek, J. Ingla-Aynes, and A. Kaverzin for useful discussions. This work was realized using NanoLabNL (NanoNed) facilities and is a part of the 'Functional Materials' programme (project number 729.002.001), financed by the Netherlands Organisation for Scientific Research (NWO). S. Chen acknowledges funding support from the European Union Horizon 2020 research and innovation programme under grant agreement No 696656 and the Spinoza Prize awarded to B. J. van Wees by NWO.

%
	



\end{document}